\documentclass[twocolumn,english,prl,superscriptaddress]{revtex4}
\usepackage[T1]{fontenc}
\usepackage[latin9]{inputenc}
\usepackage{color}
\usepackage{amsmath}
\usepackage{amssymb}
\usepackage{graphicx}
\usepackage{esint}

\makeatletter
\@ifundefined{textcolor}{}
{%
 \definecolor{BLACK}{gray}{0}
 \definecolor{WHITE}{gray}{1}
 \definecolor{RED}{rgb}{1,0,0}
 \definecolor{GREEN}{rgb}{0,1,0}
 \definecolor{BLUE}{rgb}{0,0,1}
 \definecolor{CYAN}{cmyk}{1,0,0,0}
 \definecolor{MAGENTA}{cmyk}{0,1,0,0}
 \definecolor{YELLOW}{cmyk}{0,0,1,0}
}

\@ifundefined{definecolor}
 {\usepackage{color}}{}
\@ifundefined{definecolor}{\usepackage{color}}{}
\@ifundefined{definecolor}{\usepackage{color}}{}
\@ifundefined{definecolor}{\usepackage{color}}{}

\usepackage{babel}

\usepackage{babel}

\makeatother

\usepackage{babel}
\begin{document}

\title{Floquet Fractional Chern Insulator in Doped Graphene}

\author{Adolfo G. Grushin}
\affiliation{Max-Planck-Institut f\"{u}r Physik komplexer Systeme, 01187 Dresden, Germany}
\affiliation{Instituto de Ciencia de Materiales de Madrid, CSIC, Cantoblanco,
E-28049 Madrid, Spain}
\author{\'Alvaro G\'omez-Le\'on}
\affiliation{Instituto de Ciencia de Materiales de Madrid, CSIC, Cantoblanco,
E-28049 Madrid, Spain}
\author{Titus Neupert}
\affiliation{Princeton Center for Theoretical Science, Princeton University, Princeton,
New Jersey 08544, USA}

\date{\today}

\begin{abstract}

Fractional Chern insulators are theoretically predicted states of electronic matter with emergent
topological order. They exhibit the same universal properties as the
fractional quantum Hall effect, but dispose of the need to apply a
strong magnetic field. However, despite intense theoretical
work, an experimental realization for these exotic states of matter
is still lacking. Here we show that doped graphene turns into a fractional
Chern insulator, when irradiated with high-intensity circularly polarized
light. We derive the effective steady state band structure of light-driven
graphene using Floquet theory and subsequently study the interacting
system with exact numerical diagonalization. The fractional Chern
insulator state equivalent to the $1/3$ Laughlin state appears
at $7/12$ total filling of the honeycomb lattice ($1/6$ filling of
the upper band). The state also features spontaneous ferromagnetism
and is thus an example of the spontaneous breaking of a continuous
symmetry along with a topological phase transition.

\end{abstract}

\maketitle

\begin{figure}[t]
\includegraphics[scale=0.55]{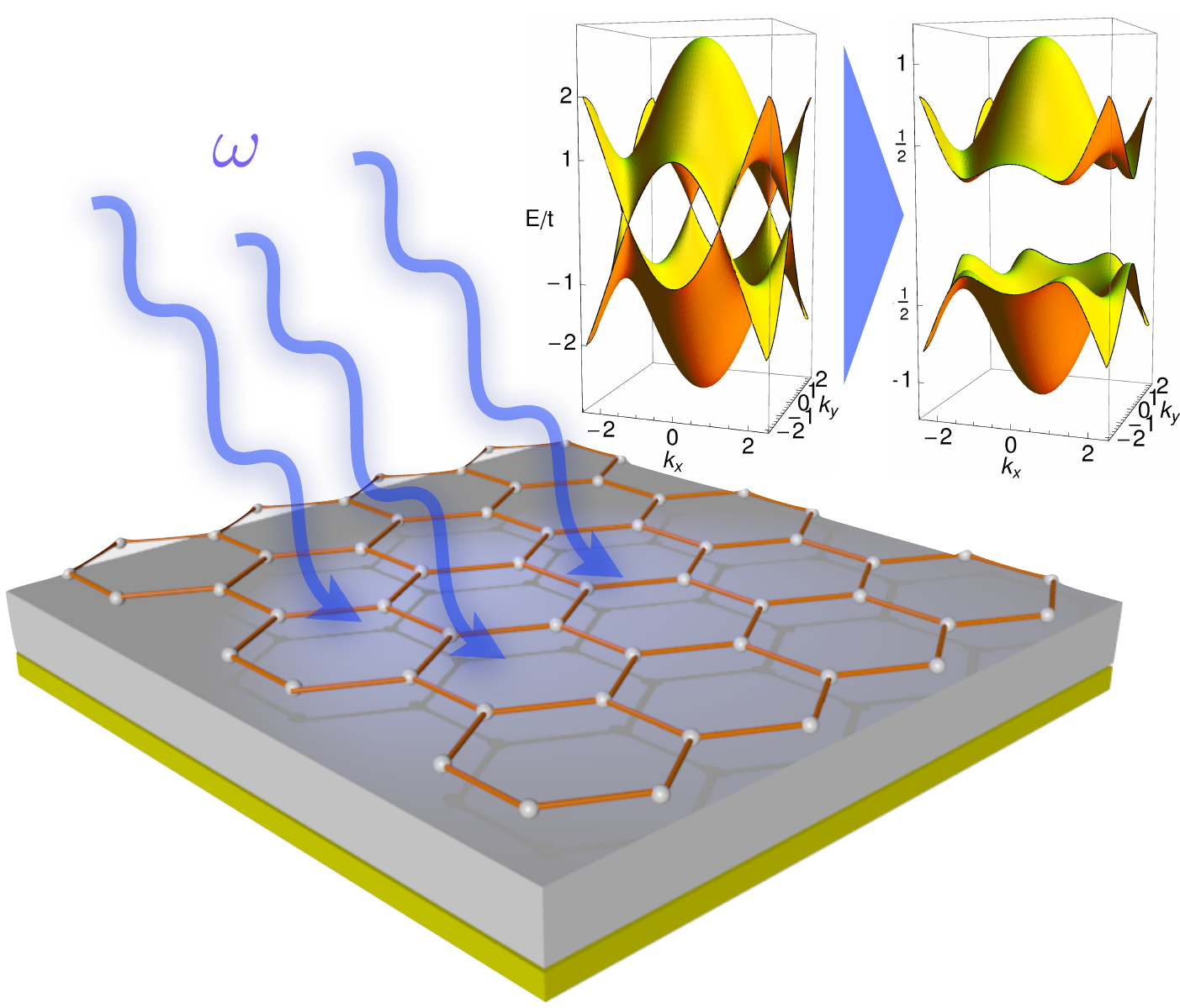} 
\caption{\label{fig:Schematic} 
Sketch of the proposed experiment. 
A graphene-flake is irradiated with light of frequency $\omega$ while a gate voltage is applied via a backgate (yellow) to change the band filling.  
In the high-frequency regime, the incident light changes the single-particle band structure into an effective Floquet band structure that acquired a gap at the Dirac points (shown are the Floquet bands for the electric field configuration $A_{x}=A_{y}=1.7$, $\phi=\pi/2$, and $\omega=10t_{1}$).
}
\end{figure}
Fractional Chern insulators (FCIs)~\cite{NSCM11,SGS11,RB11,BL13,PRS13} have been discovered numerically
in lattice models of two-dimensional electronic systems. They emerge
as the ground state of repulsively interacting electrons that partially
fill Bloch bands with a nontrivial topological attribute, namely a
nonvanishing Chern number \cite{H88}. The time-reversal symmetry (TRS)
breaking electronic hopping integrals on the lattice, that are responsible
for the band topology, take the role played by the strong magnetic
field in the factional quantum Hall effect. FCIs show that fractional
quantum Hall states can appear more generically than previously assumed,
and do not rely on the specific energetical or analytical properties
of Landau levels.

For a system to support a FCI ground state, the energetics have to
satisfy specific conditions. For example, if the topological band
is spectrally flat~\cite{NSCM11,TMW11,SGK11} or the energy scale
of the repulsive interaction exceeds the energy scales of the band~\cite{KVD12},
FCIs are favored. This is why the experimental discovery of FCIs is
still a formidable experimental challenge, despite the recent experimental
realization of its ``noninteracting'' parent band structure, the
Chern insulator or anomalous quantum Hall effect~\cite{CZF13}. Needed
are systems with a large amount of tunability, to meet both the topological
and energetical requirements. Ultracold atomic gases in optical lattices~\cite{UJM13},
artificial graphene~\cite{GMK12}, photonic crystals~\cite{RZJ13},
and light-driven solid state systems~\cite{LRG11,KOB11,DGP13b}
are such tunable platforms, all of which have been shown to potentially
host topological band structures.

In this study, we shall focus on light-driven graphene {[}see Fig.~\ref{fig:Schematic}{]}, for which
several works have proved that circularly polarized light allows to
open a gap in the Dirac cones, leading to a topologically nontrivial
state, characterized by chiral edge states~\cite{KOB11,GFA11,OA09}.
The resulting periodically driven steady state is theoretically described
using Floquet theory and is thus called a Floquet Chern insulator.
The key ingredients for its emergence are (i) the critical nature
of the Dirac electrons in graphene and (ii) the time-reversal symmetry
breaking provided by the non-linearly polarized light~\cite{DGP13a}.

Here, we show by means of numerical exact diagonalization that graphene
at $7/12$ total filling of the $\pi$-bands, when irradiated with
high-intensity circularly polarized light, realizes a ferromagnetic
FCI steady state, which we call a Floquet fractional Chern insulator
(FFCI). The FFCI state is characterized by a three-fold topological groundstate degeneracy
and a contribution to the Hall conductivity of $\sigma_{H}=\frac{1}{3}\frac{e^{2}}{h}$.
Furthermore we prove that the full SU(2) spin-rotation symmetry of the model Hamiltonian (the light-field does not couple to the
spin and the spin-orbit coupling is negligible) is spontaneously broken
for a ferromagnetic steady state, and gapless magnon excitations emerge,
coexisting with the FFCI groundstate. We emphasize that we do not
rely on a mean-field approximation to obtain this result %
\footnote{In previous work~\cite{VKV12}, a mean-field treatment was employed
to obtain a magnetically ordered background and the resulting band
structure was studied at partial filling, where the mean-field approximation
is not justified anymore.%
}.

\paragraph{Floquet approach for ac driven graphene}

We model irradiated monolayer graphene by considering spinful electrons
that populate a honeycomb lattice $\Lambda$ and interact repulsively
via their on-site ($U$) and nearest neighbor ($V$) electronic densities
$n_{i,\sigma}$, for $i\in\Lambda$ and $\sigma=\uparrow,\downarrow$,
\begin{equation}
\begin{split}H\left(\tau\right):= & H_{0}\left(\tau\right)+H_{\mathrm{int}},\\
H_{\mathrm{int}}:= & U\sum_{i}n_{i,\uparrow}n_{i,\downarrow}+V\sum_{\langle i,j\rangle}\sum_{\sigma,\sigma^{\prime}}n_{i,\sigma}n_{j,\sigma^{\prime}}.
\end{split}
\label{eq:HInitial}
\end{equation}
 For the single particle Hamiltonian $H_{0}\left(\tau\right)$, we
adopt the convention used in \cite{NSCM11,WBR12}. The time dependence
is induced via the electromagnetic vector potential $\mathbf{A}\left(\tau,\phi\right)=\left(A_{x}\sin\left(\omega\tau\right),A_{y}\sin\left(\omega\tau+\phi\right),0\right)^{\mathsf{T}}$
of the external light field, where $\omega$ is the frequency of the
light, $\phi$ is the phase difference, $A_{i}=e\mathcal{E}_{i}a/\omega m_{e}$, 
$e$ is the electron
charge, $m_{e}$ its mass, $a$ the lattice spacing, and $\mathcal{E}_{i}$
the $i$th component of the electric field. Only the single particle
Hamiltonian $H_{0}\left(\tau\right)$ is affected by this time dependence,
and not the interaction Hamiltonian. The latter remains as in the
undriven case since it is expressed in terms of the density operator.

Floquet theory provides a powerful formalism to study periodically
driven systems. It allows to easily obtain effective time evolution
operators, especially in the regime where the driving frequency is
the dominant energy scale \cite{GP13}. The single particle Hamiltonian,
which enters the time dependent Schr\"{o}dinger equation, can be obtained
following Ref. \cite{DGP13a} and the Supplementary Material, and
has the Fourier decomposition
\begin{equation}
H_{0,{\bf k}}^{q}  =  \left(\begin{array}{cc}
0 & \left(\rho_{{\bf k}}^{-q}\right)^{*}\\
\rho_{{\bf k}}^{q} & 0
\end{array}\right),\quad \rho_{\mathbf{k}}^{q}=\sum_{j}t_{j,q}^{\text{F}}e^{i\mathbf{k}\cdot\mathbf{a}_{j}},
\end{equation}
where $q\in\mathbb{Z}$ labels the Fourier component in frequency space, and the hopping integrals are given by $t_{1,q}^{\text{F}}=t_{1}J_{q}\left(A_{y}\right)e^{iq\phi}$,
$t_{2,q}^{\text{F}}=t_{1}J_{q}\left(A_{+}\right)e^{iq\Psi_{+}}$, and $t_{3,q}^{\text{F}}=t_{1}J_{-q}\left(A_{-}\right)e^{-iq\Psi_{-}}$.
Here, $t_{1}$ is the nearest-neighbor hopping integral for electrons
on the honeycomb lattice and $J_{q}\left(A\right)$ denotes the Bessel
functions of the first kind. The arguments of the Bessel functions
contain the explicit electric field configuration $A_{\pm}=\sqrt{\frac{3A_{x}^{2}}{4}+\frac{A_{y}^{2}}{4}\pm\frac{\sqrt{3}}{2}A_{x}A_{y}\cos\left(\phi\right)}$,
and the phase factors are given by $\Psi_{\pm}=\arctan\left(\frac{A_{y}\sin\left(\phi\right)}{\sqrt{3}A_{x}\pm A_{y}\cos\left(\phi\right)}\right)$. Finally, we define $\mathbf{a}_{1}=\left(0,0\right)$, $\mathbf{a}_{2}=\left(\sqrt{3},3\right)/2$,
and $\mathbf{a}_{3}=\left(-\sqrt{3},3\right)/2$ as the unit cell vectors of the honeycomb lattice.

\begin{figure}[t]
\includegraphics[scale=0.5]{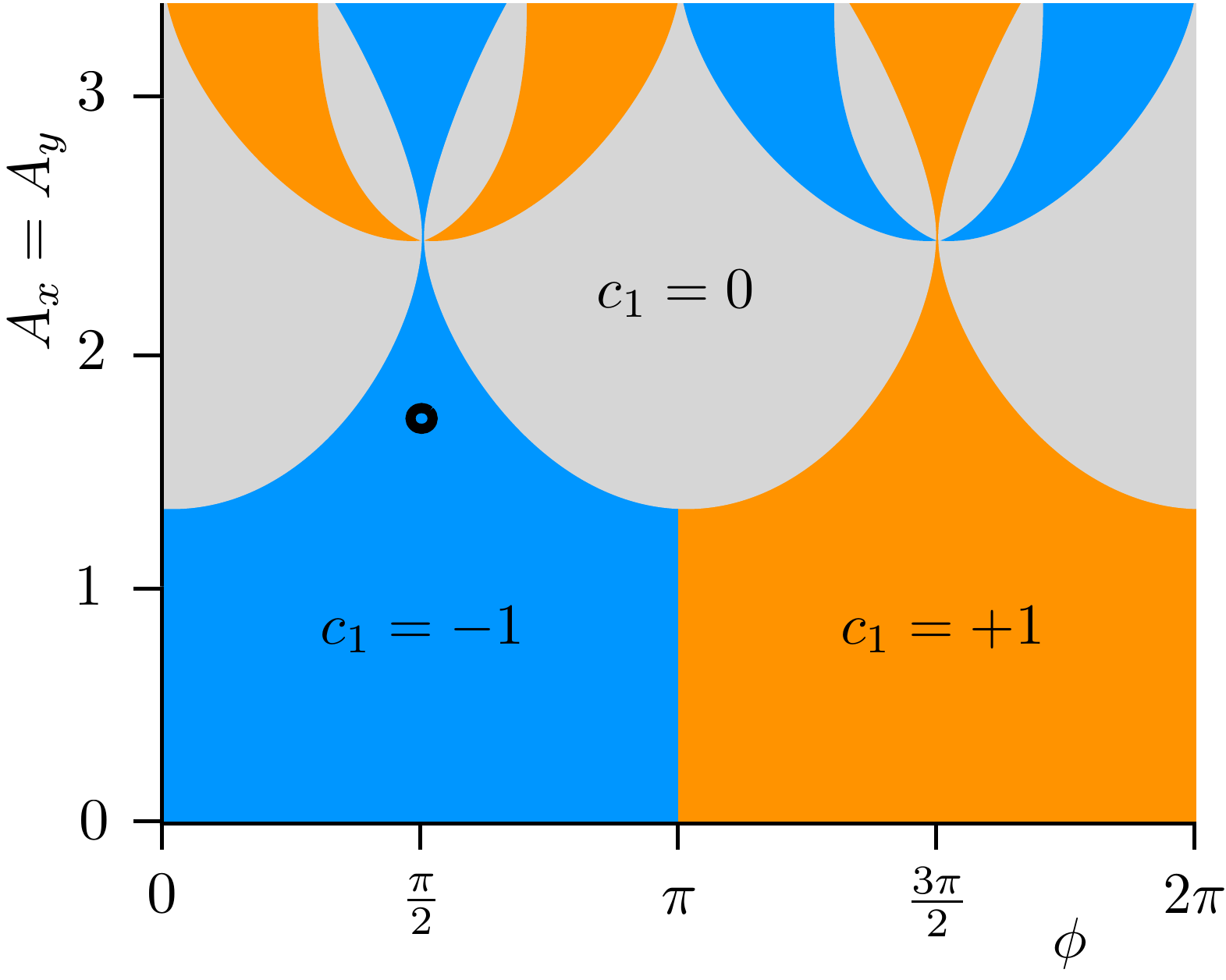} \caption{\label{fig: Chern number phase diagram} Topological phase diagram
of periodically driven graphene at high frequency regime ($\omega=10t_{1}$)
for circularly polarized light. The phase difference $\phi$ and the
field amplitude $A_{x}=A_{y}$ allow to tune the Chern number of the
lower spin-degenerate bands of the effective Floquet-Bloch Hamiltonian~\eqref{eq: Floquet-Bloch}
between $c_{1}=-1$ and $c_{1}=+1$. The black dot indicates the parameter
values for which exact diagonalization calculations are presented
in Figs.~\ref{fig: Spectrum} and~\ref{fig: Flux Tower}.}
\end{figure}

In this work we focus on the high frequency regime ($\omega\gg t_{1}$),
where the system dynamics can be accurately described by a static
effective Hamiltonian, which can be expanded order by order in $t_{1}/\omega$.
To zeroth order in $t_{1}/\omega$, the hopping integrals are renormalized
by zeroth order Bessel functions only, with no breaking of TRS. To
first order in $t_{1}/\omega$, corrections to the effective Hamiltonian
must be considered due to hybridization with the nearest Floquet bands.
Importantly, for non linear field polarization, TRS is broken by these
corrections, a prerequisite to realize a Floquet Chern insulator.
The effective Hamiltonian for long-time dynamics is then defined by
including higher Fourier components of the Hamiltonian, leading to
the $2\times2$ Floquet-Bloch Hamiltonian (for details see the Supplementary
Material): 
\begin{equation}
\begin{split}H_{\text{eff},{\bf k}}=H_{0,\mathbf{k}}^{0}-\frac{1}{\omega}\Bigl( & \left[H_{0,{\bf k}}^{0},H_{0,{\bf k}}^{-1}\right]\\
 & -\left[H_{0,{\bf k}}^{0},H_{0,{\bf k}}^{1}\right]+\left[H_{0,{\bf k}}^{-1},H_{0,{\bf k}}^{1}\right]\Bigr).
\end{split}
\label{eq: Floquet-Bloch}
\end{equation}
 Then, the effective single particle Hamiltonian expressed using the second-quantized  operators $c_{{\bf k},\sigma}^{\dagger}=(c_{{\bf k},\sigma,A}^{\dagger},c_{{\bf k},\sigma,B}^{\dagger})$, that create an electron with momentum $\bf{k}$ and spin $\sigma$ in sublattice $A$ and $B$, respectively, reads 
\begin{equation}
H_{\text{eff},\mathbf{k}}:=\sum_{{\bf k}\in\mathrm{BZ}}\sum_{\sigma=\uparrow,\downarrow}c_{{\bf k},\sigma}^{\dagger}H_{\text{eff},{\bf k}}c_{{\bf k},\sigma}.\label{eq: effective noninteracting Hamiltonian}
\end{equation}
 The Hamiltonian \eqref{eq: effective noninteracting Hamiltonian}
has two pairs of spin-degenerate bands that touch in two Dirac points
for $t_{1}/\omega\to0$. The correction to first order in $t_{1}/\omega$
in the $2\times2$ Floquet-Bloch Hamiltonian $H_{\text{eff},{\bf k}}$
is proportional to the third Pauli matrix $\sigma_{z}$. Due to the TRS breaking it can thus
potentially open a Haldane-type gap \cite{H88} in the spectrum, so
that the resulting spin-degenerate bands can acquire a Chern number $c_{1}=\pm1$
for each spin species. The phase diagram for different externally
tunable parameters is shown in Fig.~\ref{fig: Chern number phase diagram}.

 It is worth emphasizing that in the high frequency limit relevant
for this work, the Hamiltonian \eqref{eq: effective noninteracting Hamiltonian}
is a time independent effective Hamiltonian that governs stroboscopic
evolution. Therefore, it allows to fill the bands as in the case of
time independent systems. 

\paragraph{Exact diagonalization results}

We are now going to show that the ground state of the Floquet Hamiltonian
$H_{\mathrm{eff}}+H_{\mathrm{int}}$, and with this the steady state
of the driven Hamiltonian \eqref{eq:HInitial}, can be tuned into
a FFCI by controlling (i) the filling of the system with electrons
and (ii) the amplitude and phase of the light field. The emergence
of a FFCI depends crucially on the ratios between the energy scales
of the single-particle band gap $\Delta$, the single particle band
width $W$ and the repulsive electron-electron interactions, 
where mathematical band flatness is not always the optimal choice \cite{GNC12}.
In graphene, the interaction parameters are given by $U=3t$ and $V=2t$ \cite{WSF11}.
Choosing incident light with amplitude $A_{x}=A_{y}=1.7$ and the phase
shift $\phi=\pi/2$, for example, results in 
an $H_{\mathrm{eff}}$ with $\Delta/W=0.6$. We study
the system above half-filling, so that the upper spin-degenerate band
of $H_{\mathrm{eff}}$ is partially filled. Given the size of the single particle gap $\Delta$, it is reasonable 
to approximate the states in the lower
band to be occupied with probability one, even in the interacting
many-body ground state~\cite{KVD12}.
We thus ignore those single-particle states and project the degrees of
freedom of the system to the upper spin-degenerate band of $H_{\mathrm{eff}}$
by means of the projector $P$ and to study the Hamiltonian 
\begin{equation}
H_{\mathrm{proj}}=PH_{\mathrm{eff}}P+PH_{\mathrm{int}}P.\label{eq: projected Hamiltonian}
\end{equation}
 Previous studies have shown that FCI ground states equivalent to
the $1/m$ Laughlin state of the fractional Quantum Hall effect in Landau levels emerge quite generically if a flat band
with Chern number $c_{1}=1$ is populated with spineless fermions at a filling
$\nu=1/m,\ m\in\mathbb{Z}$ \cite{NSCM11,SGS11,RB11,WBR12}. In contrast,
we are considering a \textit{dispersionful} band with Chern number $c_{1}=1$ that is
partially filled with \textit{spinful} electrons in such a way that
the Hamiltonian is SU(2) spin-rotation symmetric. In anticipation
of a spontaneous breaking of the SU(2) symmetry by the many-body ground
state, we therefore study the system at $\nu=1/(2m)$ filling to obtain
an FFCI that is equivalent to a $\nu=1/m$ Laughlin state. We have
performed numerical exact diagonalization of the Hamiltonian~\eqref{eq: projected Hamiltonian}
on lattices with $L_{x}\times L_{y}=4\times3$ and $L_{x}\times L_{y}=4\times6$
unit cells with $N=4$ and $N=8$ electrons, respectively, with periodic
boundary conditions in place. Good quantum numbers of the many-body
states are the total spin $S=0,\cdots,N/2$, the total spin-$z$ component
$S_{z}=-N/2,\cdots,N/2$ and the center of mass momentum $Q\in\left[0,L_{x}\times L_{y}-1\right]$.
All results and conclusions presented below extend to both lattice
sizes and we focus on the $L_{x}\times L_{y}=4\times6$ lattice here,
delegating the consistency check with $L_{x}\times L_{y}=4\times3$
to the Supplementary Material. \\
\begin{figure}[t]
\includegraphics[scale=0.41, page=1]{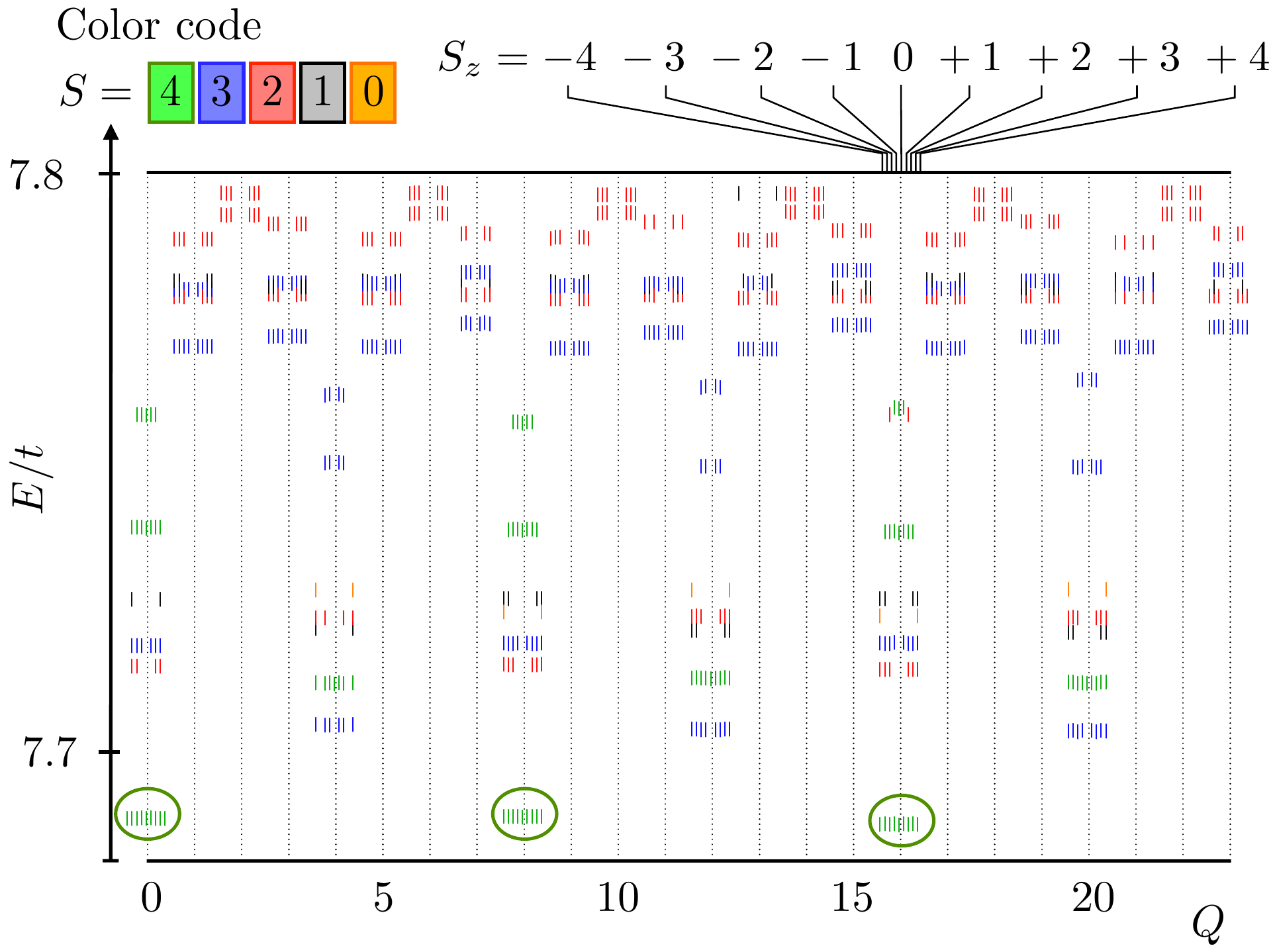} \caption{\label{fig: Spectrum} Low-lying portion of the energy spectrum of
Hamiltonian~\eqref{eq: projected Hamiltonian} on a $L_{x}\times L_{y}=4\times6$
with $N=8$ particles. Encircled is the ground state manifold with
3-fold topological (quasi-) degeneracy of the fractional Chern insulator
and $N+1=9$ fold degeneracy as a precursor of spontaneous symmetry
breaking toward a ferromagnetic phase in the thermodynamic limit.
The good quantum numbers total momentum $Q=k_{x}+L_{x}k_{y}$, total
spin $S$ and total spin in $z$ direction $S_{z}$ are indicated.}
\end{figure}
We observe that the low-energy states have both exact and approximate
degeneracies: In each of the $(N+1)$ sectors of $S_{z}$, three low-lying
states are found which are not exactly, but approximately, degenerate.
Each of these three states has an exactly degenerate partner in every
other $S_{z}$ sector (see Fig.~\ref{fig: Spectrum}). The total
ground state degeneracy that we anticipate in the thermodynamic limit
is thus $3(N+1)$. As all ground states have the maximum spin $S=N/2$,
we interpret the exact $(N+1)$-fold degeneracy as a finite-size precursor
of a spontaneous breaking of the SU(2) symmetry towards a ferromagnetic
ground state in the thermodynamic limit. Both the tower-of-states
structure~\cite{LMF10} in the spectrum as a function of $S$ {[}Fig.~\ref{fig: Flux Tower}
(b){]} and the exactness of the degeneracy support this conclusion
(the order parameter $S_{z}$ of ferromagnetism commutes with the
Hamiltonian thus rendering the ground state degeneracy exact already
for finite systems). In contrast, the 3-fold approximate degeneracy
in each $S_{z}$ sector is of topological origin. It is the $m$-fold
topological ground state degeneracy of a $1/m$ Laughlin state on
the torus in the case $m=3$. The non-local Wilson-loop order parameter
does not commute with the Hamiltonian, rendering the degeneracy approximate
in the finite system. Further supporting arguments that the three
low-lying states in each $S_{z}$ sector are indeed topologically
ordered FFCI states are: (i) By inserting a flux in the torus (which
is equivalent to changing the boundary conditions from periodic to
twisted~\cite{NTDW85}), the three states permute and return to their
original order after three flux quanta {[}see Fig.~\ref{fig: Flux Tower}
(a){]}. This evidences charge fractionalization with quasiparticles
of charge $e/3$ and indicates that the topological groundstates survive in the thermodynamic limit.
(ii) The three states occur
at the momentum sectors $Q$ that are predicted by the counting rule
of Ref.~\cite{RB11} that is based on a $1/3$ Laughlin state. (iii)
Any superposition of the three ground states has a nearly constant
charge density in position space, which excludes that these states
would form a charge-density wave in the thermodynamic limit (see Supplementary
Material).

\begin{figure}[t]
\includegraphics[scale=0.41,page=2]{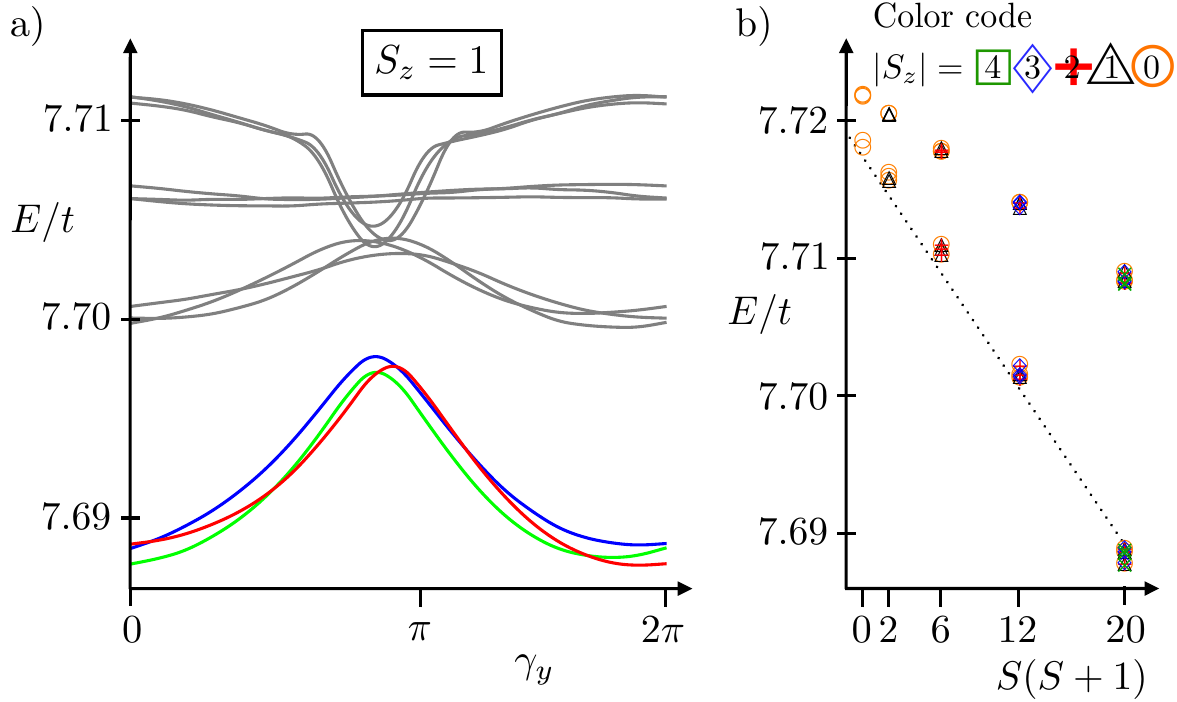} \caption{\label{fig: Flux Tower} a) Spectral evolution of the energy spectrum
of Fig.~\ref{fig: Spectrum} in the sector with $S_{z}=1$ upon inserting
a flux $\gamma_{y}$ into the system, which is synonymous to twisting
the boundary conditions in $y$ direction with a complex phase $e^{\mathrm{i}\gamma_{y}}$.
The three fractional Chern insulator ground states evolve independently
of the rest of the spectrum, and trade places, which signals their
topological degeneracy and the charge fractionalization. b) The energy
spectrum of Fig.~~\ref{fig: Spectrum} plotted against the total
spin $S(S+1)$ reveals the tower of states which evidencing the ferromagnetic
nature of the ground state.}
\end{figure}

From the presented results we conclude that the steady state of graphene
at $7/12$ total filling of the $\pi$ bands (or $1/6$ filling of
the band above the Dirac cone) is a ferromagnetic FFCI. This distinct
driven topological state has gapped charged excitations but supports gapless
spin excitations, namely ferromagnetic magnons. The state features both conventional 
order and gapless topological order. 
Experimentally, its signature is a fractional contribution of 
$\sigma_{H}=\frac{1}{3}\frac{e^2}{h}$ (in addition to the
integer contribution of the lower band) defined for the
driven system \cite{KS13}.

 To certify the robustness of the FFCI state we have also tested
its stability under changing some of the conditions discussed above.
Firstly, we have investigated its fate upon changing to different
light polarizations (using $A_{x}/A_{y}=\sqrt{3},1/\sqrt{3}$ as two examples). The
FFCI is still the groundstate of the system as long the Chern number
of the non interacting band is $c_{1}=\pm1$. Secondly, we have investigated
its appearance on a different lattice system. In particular, we find
that for spinless electrons on the $\pi$-flux square lattice with
nearest neighbor repulsive interaction the ground state is also the FFCI whenever $(A_{x}=A_{y},\phi)$
are such that the non-interacting bands have $c_{1}=\pm1$ (see Supplementary
Material). Both of these results evidence the ubiquitousness and robustness 
of the FFCI state.

\paragraph{Experimental realization:}

The practical realization of this novel state in graphene possess
two experimental challenges. The first is to reach incident field
amplitudes $\mathcal{E}_{x}=\mathcal{E}_{y}\sim\omega\gg t_{1}$.
Although it is in principle possible to reach such a regime, today experiments
have only explored amplitudes of one order of magnitude lower \cite{TBT12}.
However, given the robustness and tuneability of the effect it is conceivable 
that lower frequencies and field amplitudes can in fact be sufficient to access the FFCI 
state in the thermodynamic limit. 
The second experimental issue is reaching
the necessary filling factor of graphene's band structure. In particular,
the electron density at $1/6$ filling of the upper band is of the
order of $1\times10^{14}e/cm^{2}$, still below the van Hove singularity
in graphene. Even higher values up to the van Hove singularity have
already been reached by chemical doping~\cite{CBO10} and there are
promising other routes by using for example polymer electrolytes~\cite{PJA10}. Finally,
other 2D materials with Dirac electrons such as silicene can potentially be used to tune
the band structure parameters and host similar phases.\\

To summarize, we have found that doped graphene, irradiated with 
polarized light, undergoes an interaction driven spontaneous topological phase transition 
to a Floquet fractional Chern insulator (FFCI) state that features both topological and spontaneous ferromagnetic order. 
The robustness and tuneability of the FFCI state, appearing for different parameters, polarizations and lattices,
evidences that this phase can be experimentally discovered in graphene.
More generally, our work opens up a promising route to achieve tuneable realizations of elusive interacting  fermionic and bosonic phases by periodically driving interacting systems.
\paragraph{Acknowledgements:} We acknowledge E. Berg, D. Podolsky, G. Platero, M.A.H. Vozmediano, for useful discussions. 
Financial support from PIB2010BZ-00512 (A.G.G.),  JAE program,  MAT 2011-24331 and ITN, grant 234970 (EU) (A.G-L.) and
Swiss national science foundation (T.N.) is greatly acknowledged.


\begin{widetext}

\appendix

\section{Explicit calculation of the Floquet Hamiltonian}

We initially consider the full time dependent Hamiltonian 
\begin{equation}
H\left(\tau\right)=H_{0}\left(\tau\right)+U\sum_{i}n_{\uparrow,i}n_{\downarrow,i}+V\sum_{\langle i,j\rangle}\sum_{\sigma,\sigma^{\prime}}n_{i,\sigma}n_{j,\sigma^{\prime}},\label{eq:HInitial-1}
\end{equation}
 where
\begin{eqnarray*}
H_{0}\left(\tau\right) & = & \left(\begin{array}{cc}
0 & \rho\left(\mathbf{k},\tau\right)^{*}\\
\rho\left(\mathbf{k},\tau\right) & 0
\end{array}\right),\ \rho\left(\mathbf{k},\tau\right)=\sum_{j=1}^{3}t_{j}\left(\tau\right)e^{i\mathbf{k}\cdot\mathbf{a}_{j}},
\end{eqnarray*}
and $t_{j}\left(\tau\right)=t_{1}e^{i\mathbf{A}\left(\tau\right)\cdot\mathbf{d}_{j}}$,
which is the usual coupling times a phase factor due to the presence
of a vector potential. The vectors $\mathbf{a}_{j}$ are the unit cell vectors
$\mathbf{a}_{1}=a\left(0,0\right)$, $\mathbf{a}_{2}=\left(\sqrt{3},3\right)a/2$,
$\mathbf{a}_{3}=\left(-\sqrt{3},3\right)a/2$, and the distance to
the nearest neighbors $\mathbf{d}_{j}$ are given by $\mathbf{d}_{1}=a\left(0,-1\right)$,
$\mathbf{d}_{2}=a\left(\sqrt{3},1\right)/2$, and $\mathbf{d}_{3}=a\left(-\sqrt{3},1\right)/2$.
Importantly, the interaction terms does not couple to the ac field,
which can be understood as follows: The interaction terms depend
on the momentum difference between the electrons, and then for a spatially
homogeneous fields this is time independent. Alternatively, as the
coupling with the electric field can be also writen as $\vec{\mathcal{E}}\left(t\right)\cdot\vec{x}$,
where the electric field $\vec{\mathcal{E}}\left(t\right)=-\partial_{t}\mathbf{A}\left(t\right)$,
the field will only couple to those terms which does not commute with
the position operator. Therefore, on site interactions are proportional
to the possition operator and will not become time dependent.\\
Following Ref. \cite{DGP13a} the Floquet operator can be calculated in Sambe space, leading to 
\begin{eqnarray*}
\mathcal{H}{}_{0,\mathbf{k}}^{q} & = & \left(\begin{array}{cc}
0 & \left(\rho_{\mathbf{k}}^{-q}\right)^{*}\\
\rho_{\mathbf{k}}^{q} & 0
\end{array}\right)-n\omega\delta_{n,m},\ \rho_{\mathbf{k}}^{q}=\sum_{j}t_{j,q}^{F}e^{i\mathbf{k}\cdot\mathbf{a}_{j}},
\end{eqnarray*}
 where the hoppings are 
\begin{eqnarray*}
t_{1,q}^{F} & = & t_{1}J_{q}\left(A_{y}\right)e^{iq\phi},\ t_{2,q}^{F}=t_{1}J_{q}\left(A_{+}\right)e^{iq\Psi_{+}},\ \text{and }t_{3,q}^{F}=t_{1}J_{-q}\left(A_{-}\right)e^{-iq\Psi_{-}},
\end{eqnarray*}
with $q\equiv n-m$, being $n$ and $m$ the Fourier components of the Fourier expansion of each Floquet state.  The arguments of the Bessel functions and phase factors are
\begin{eqnarray*}
A_{\pm} & = & \sqrt{\frac{3A_{x}^{2}}{4}+\frac{A_{y}^{2}}{4}\pm\frac{\sqrt{3}}{2}A_{x}A_{y}\cos\left(\phi\right)},\ \Psi_{\pm}=\arctan\left(\frac{A_{y}\sin\left(\phi\right)}{\sqrt{3}A_{x}\pm A_{y}\cos\left(\phi\right)}\right).
\end{eqnarray*} 
\[
\]
 To obtain the effective Hamiltonian let us consider a stroboscopic
evolution operator over a period $T$ 
\[
U\left(T\right)=\mathcal{T}e^{-i\int_{0}^{T}H_{0}\left(\tau\right)d\tau}\simeq e^{-iH_{eff}T}.
\]
 Next, due to its time periodicity we consider the Fourier decomposition
of the Hamiltonian 
\begin{equation}
H_{0}\left(\tau\right)=\sum_{n=-\infty}^{\infty}H_{0}^{n}e^{in\omega\tau}\simeq H_{0}^{0}+H_{0}^{1}e^{i\omega\tau}+H_{0}^{-1}e^{-i\omega\tau},\label{eq:FirstOrderHam1}
\end{equation}
 where we have considered just the first harmonic contribution, and
$H_{0}^{\pm1}=\frac{1}{T}\int_{0}^{T}H_{0}\left(\tau\right)e^{\mp i\omega\tau}d\tau$.
Expanding the matrix exponential in Taylor series we obtain: 
\[
e^{-i\int_{0}^{T}H_{0}\left(\tau\right)d\tau}\simeq1-i\int_{0}^{T}H_{0}\left(\tau\right)d\tau+\frac{\left(-i\right)^{2}}{2}\int_{0}^{T}H_{0}\left(\tau_{1}\right)d\tau_{1}\int_{0}^{T}H_{0}\left(\tau_{2}\right)d\tau_{2}+\ldots
\]
 Then, in terms of the previous expansion, the stroboscopic evolution
operator is given by 
\begin{eqnarray*}
U\left(T\right) & \simeq & \mathcal{T}\left\{ 1-i\int_{0}^{T}H_{0}\left(\tau\right)d\tau+\frac{\left(-i\right)^{2}}{2}\int_{0}^{T}H_{0}\left(\tau_{1}\right)d\tau_{1}\int_{0}^{T}H_{0}\left(\tau_{2}\right)d\tau_{2}\right\} \\
 & = & 1-i\int_{0}^{T}H_{0}\left(\tau\right)d\tau-\frac{1}{2}\left[\int_{0}^{T}d\tau_{1}\int_{0}^{\tau_{1}}d\tau_{2}H_{0}\left(\tau_{1}\right)H_{0}\left(\tau_{2}\right)+\int_{0}^{T}d\tau_{2}\int_{0}^{\tau_{2}}d\tau_{1}H_{0}\left(\tau_{2}\right)H_{0}\left(\tau_{1}\right)\right],
\end{eqnarray*}
 where in the last line we have applied the time ordering operator
$\mathcal{T}$. Using Eq.~\eqref{eq:FirstOrderHam1} we can perform the
time integrals, and reordering terms compare with the expansion of
the effective time evolution operator 
\begin{eqnarray*}
U\left(T\right) & \simeq & 1-iH_{0}^{0}T-\frac{T}{\omega}\left\{ \pi\left(H_{0}^{0}\right)^{2}-i\left(\left[H_{0}^{0},H_{0}^{-1}\right]-\left[H_{0}^{0},H_{0}^{1}\right]+\left[H_{0}^{-1},H_{0}^{1}\right]\right)\right\} \\
 & \simeq & 1-iH_{\text{eff}}T-\frac{1}{2}H_{\text{eff}}^{2}T^{2}+\ldots
\end{eqnarray*}
 The effective Hamiltonian is finally given by 
\begin{equation}
H_{\text{eff}}=H_{0}^{0}-\frac{1}{\omega}\left(\left[H_{0}^{0},H_{0}^{-1}\right]-\left[H_{0}^{0},H_{0}^{1}\right]+\left[H_{0}^{-1},H_{0}^{1}\right]\right),\label{eq:Heff}
\end{equation}
 which is eq.~\eqref{eq: Floquet-Bloch} of the main text. Explicitly, the single particle Hamiltonian in $\mathbf{k}$
domain $H_{\text{eff},\mathbf{k}}=t_{1}\vec{g}\left(\mathbf{k}\right)\cdot\vec{\sigma}$
is given by 
\begin{align*}
g_{x} & =-J_{0}\left(A_{-}\right)\cos\left(\frac{\sqrt{3}}{2}k_{x}-\frac{3}{2}k_{y}\right)+J_{0}\left(A_{+}\right)\cos\left(\frac{\sqrt{3}}{2}k_{x}+\frac{3}{2}k_{y}\right)+J_{0}\left(A_{y}\right),\\
g_{y} & =J_{0}\left(A_{-}\right)\sin\left(\frac{\sqrt{3}}{2}k_{x}-\frac{3}{2}k_{y}\right)-J_{0}\left(A_{+}\right)\sin\left(\frac{\sqrt{3}}{2}k_{x}+\frac{3}{2}k_{y}\right),\\
g_{z} & =-\frac{4t_{1}}{\omega}\{J_{0}\left(A_{+}\right)\left[J_{1}\left(A_{-}\right)\cos\left(\sqrt{3}k_{x}\right)\cos\left(\Psi_{-}\right)-J_{1}\left(A_{+}\right)\cos\left(\Psi_{+}\right)+J_{1}\left(A_{y}\right)\cos\left(\phi\right)\cos\left(\frac{\sqrt{3}}{2}k_{x}+\frac{3}{2}k_{y}\right)\right]\\
 & +J_{0}\left(A_{-}\right)\left[J_{1}\left(A_{-}\right)\cos\left(\Psi_{-}\right)-J_{1}\left(A_{+}\right)\cos\left(\sqrt{3}k_{x}\right)\cos\left(\Psi_{+}\right)+J_{1}\left(A_{y}\right)\cos\left(\phi\right)\cos\left(\frac{\sqrt{3}}{2}k_{x}-\frac{3}{2}k_{y}\right)\right]\\
 & +J_{0}\left(A_{y}\right)\left[J_{1}\left(A_{-}\right)\cos\left(\Psi_{-}\right)\cos\left(\frac{\sqrt{3}}{2}k_{x}-\frac{3}{2}k_{y}\right)-J_{1}\left(A_{+}\right)\cos\left(\Psi_{+}\right)\cos\left(\frac{\sqrt{3}}{2}k_{x}+\frac{3}{2}k_{y}\right)+J_{1}\left(A_{y}\right)\cos\left(\phi\right)\right]\\
 & -J_{1}\left(A_{y}\right)\left[J_{1}\left(A_{-}\right)\sin\left(\frac{\sqrt{3}}{2}k_{x}-\frac{3}{2}k_{y}\right)\sin\left(\Psi_{-}+\phi\right)+J_{1}\left(A_{+}\right)\sin\left(\frac{\sqrt{3}}{2}k_{x}+\frac{3}{2}k_{y}\right)\sin\left(\phi-\Psi_{+}\right)\right]\\
 & +J_{1}\left(A_{-}\right)J_{1}\left(A_{+}\right)\sin\left(\sqrt{3}k_{x}\right)\sin\left(\Psi_{-}+\Psi_{+}\right)\}.
\end{align*}
 Eq.~\eqref{eq:Heff} is the single particle Hamiltonian considered all
throughout this work. From it it is possible to extract a topological phase diagram for the Chern number $c_{1}$
as a function of the external parameters of the electric field shown in Fig.~\ref{fig: Chern number phase diagram}.
The band structure for the point $A_{x}=A_{y}=1.7$, and $\phi=\pi/2$ focus of this work is shown in Fig.~\ref{fig:Schematic}.

\begin{figure}[t]
\includegraphics[scale=0.38, page=3]{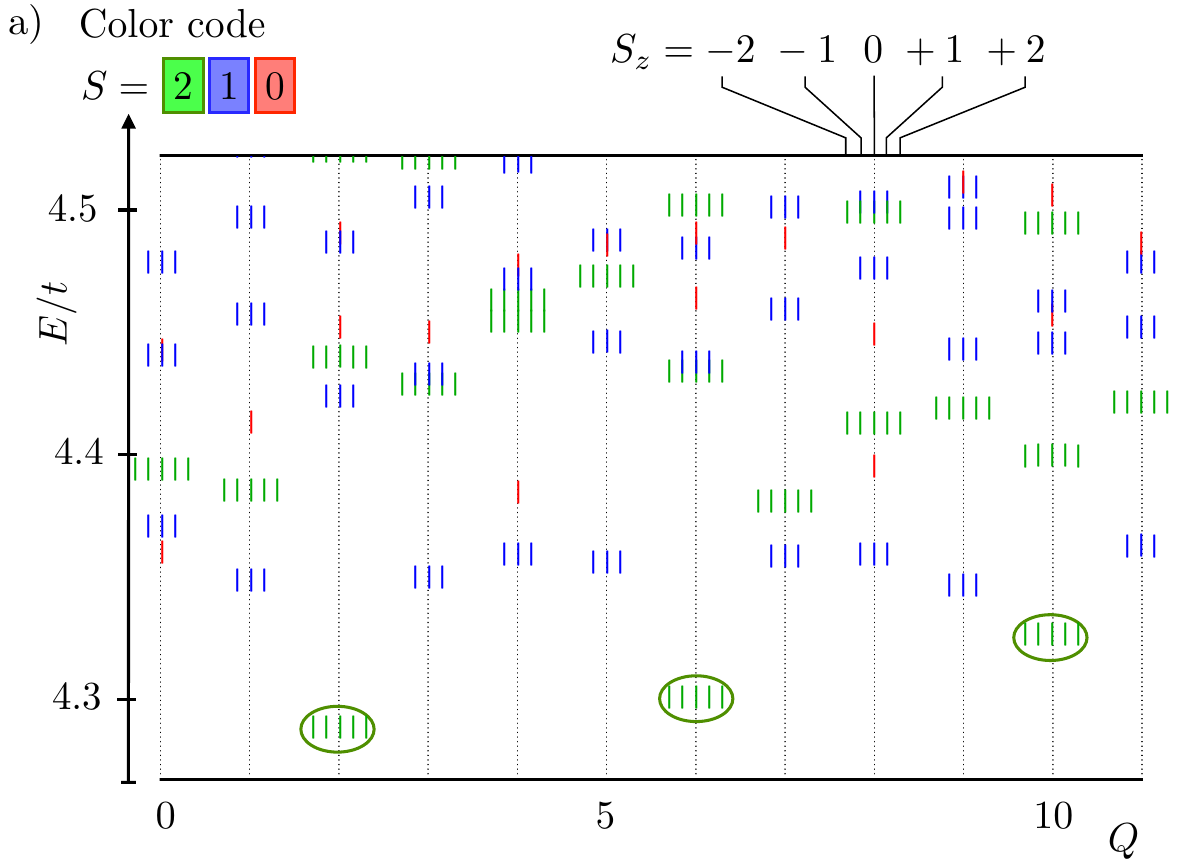} \hspace{0.2cm}
\includegraphics[scale=0.38, page=4]{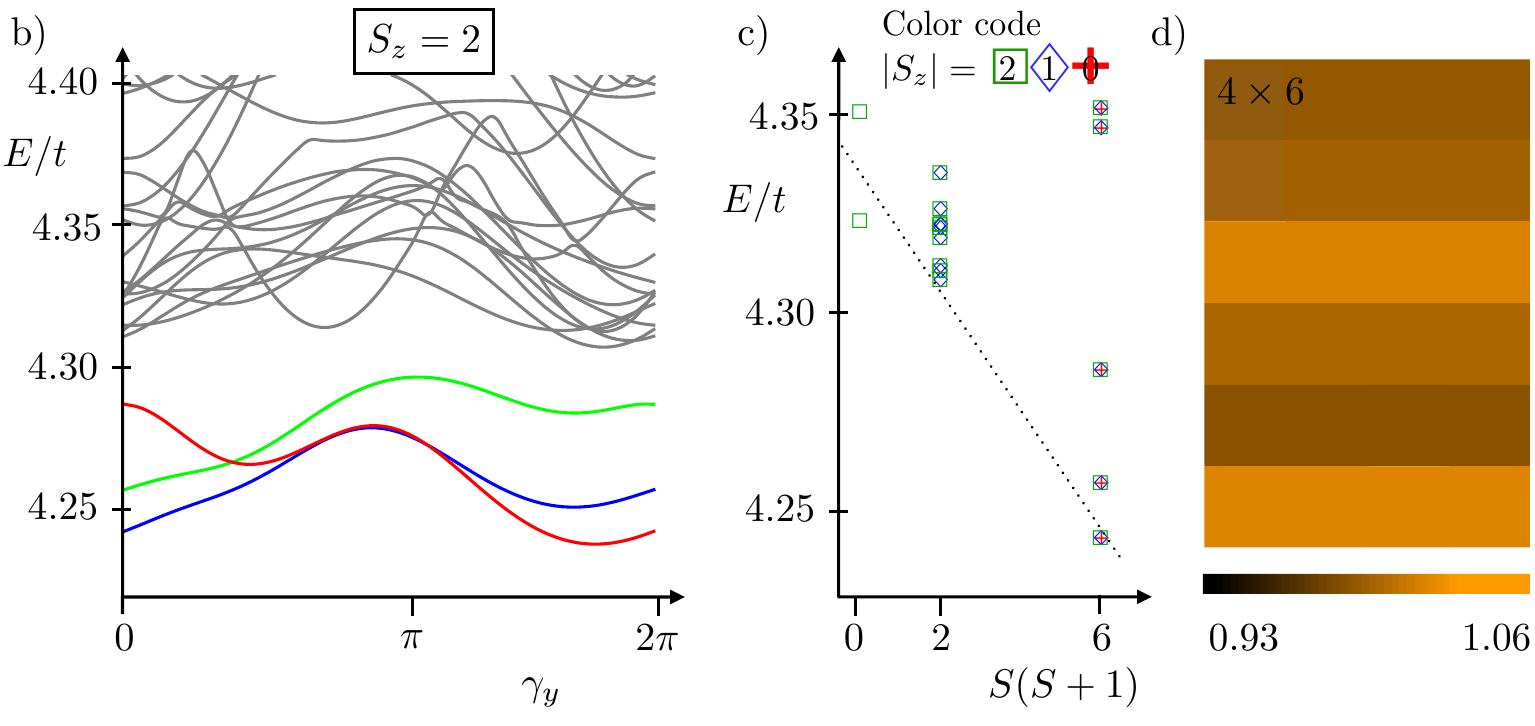} 
\caption{\label{fig:Band-structure} 
a) Many-body spectrum, b) flux insertion, and c) tower-of-states for the
same parameters for a $L=4\times3$ lattice that evidence the presence
of a ferromagnetically ordered FFCI state. d) We show also the local
density profile $n_{\boldsymbol{r}}^{(i)}$ for the case of a $L=4\times 6$ lattice
defined in \eqref{eq: density n(r)}
for one representative many-body groundstate. The small density variation
is symptomatic of the FCI state.}
\end{figure}

\section{Further numerical evidence in favor of the presence and robustness
of the Floquet fractional Chern insulator phase:}

In this section, additional numerical results mentioned in the main
text are presented as supporting evidence for the ferromagnetic Floquet fractional
Chern insulator (FFCI).

\subsection{Spectrum and Tower of states for $L=4\times3$ $N=4$}

First, we focus on a different lattice size, $L=4\times3$ with $N=4$
particles, to compare with $L=4\times6$ with $N=8$ particles discussed
in the main text. The spectrum, flux insertion and tower-of-states for this lattice
size is presented in Fig.~\ref{fig:Band-structure} (a)-(c). The figures show the same qualitative
features as for the $L=4\times6$ lattice case discussed in the main text.\\
 Firstly, as shown in Fig.~\ref{fig:Band-structure} (a) the groundstate is three-fold degenerate
for each $S_{z}$ sector. The groundstates appear at the total momenta
$Q$ predicted by the counting rule of Ref.~\cite{RB11}. Under an adiabatic flux insertion
in the $\gamma_{y}$ direction, the states interchange signalling the topological degeneracy and 
charge fractionalization {[}see Fig.~\ref{fig:Band-structure} (b){]}. Furthermore,
all groundstates have the maximum spin $S=N/2$ which is evidence
for the ferromagnetic nature of the FFCI state. This picture is further
supported by the tower-of-states of Fig.~\ref{fig:Band-structure} (c) 
that shows that the higher values
of $S(S+1)$ have the lowest energy, a finite size signature of the
thermodynamic spontaneous symmetry breaking of the SU(2) symmetry,
as discussed in the main text.

\subsection{Charge density profile:}

To ascertain the topological nature of the Floquet fractional Chern
insulator state it is important to rule out the possibility of the
appearance of a charge density wave (CDW) order, since both states
can present the same momentum counting \cite{WBR12}. A way to distinguish
between both phases was put forward in Ref.~\cite{GNC12} by calculating
the real space density profile, a procedure that we follow here. To
this extent, we consider the local density operator defined by \begin{subequations}
\begin{equation}
\rho_{\boldsymbol{r}}:=\frac{1}{L_{1}^{\,}L_{2}^{\,}}\sum_{\boldsymbol{q},\boldsymbol{k}}\sum_{s}e^{\mathrm{i}\,\boldsymbol{q}\cdot\boldsymbol{r}}\, c_{\boldsymbol{k}+\boldsymbol{q},s}^{\dagger}\, c_{\boldsymbol{k},s},
\end{equation}
 For the quasi degenerate ground state manifold $|\Psi_{1}^{\,}\rangle,\cdots,|\Psi_{n}^{\,}\rangle$
($n=3$ for the Laughlin state) we construct the matrix with elements
\begin{equation}
\varrho_{\boldsymbol{r};ij}:=\langle\Psi_{i}|\rho_{\boldsymbol{r}}|\Psi_{j}\rangle.\label{eq: rhomatrix}
\end{equation}
 \end{subequations} The next step is to obtain a set of $n$ maps
of the local fermion density in the ground-state manifold. Labelling
$v_{\rho;\boldsymbol{r}_{0}^{\,}}^{(i)},\ i=1,\cdots,n$ the set of
orthonormal eigenvectors of ${\varrho}_{\boldsymbol{r}_{0}^{\,}}^{\,}$
at some arbitrarily chosen site $\boldsymbol{r}_{0}^{\,}$ we evaluate
the $n$ real functions 
\begin{equation}
n_{\boldsymbol{r}}^{(i)}:=v_{\rho;\boldsymbol{r}_{0}^{\,}}^{(i)\dagger}\;\,{\varrho}_{\boldsymbol{r}}^{\,}\;\, v_{\rho;\boldsymbol{r}_{0}^{\,}}^{(i)}\;,\qquad i=1,\cdots,n.\label{eq: density n(r)}
\end{equation}
 The functions $n_{\boldsymbol{r}}^{(i)}$ are density maps that show
the variation of the local fermion density in position space. Featureless
or flat density maps would be strong evidence in favour an FCI. Therefore,
finding strongly patterned density maps is a signature of a ground-state
manifold that supports a CDW.\\
 A typical charge density map for the three-fold ground state manifold
of the FFCI state is shown in Fig.~\ref{fig:Band-structure}(d) for
the particular point $A_{x}=A_{y}=1.7$ $\phi=\pi/2$ discussed in
the main text. The density variation in position space is small, of
about $\sim5\%$ which is indication of a FCI state. We attribute
the small deviation to the lattice anisotropy, already observed in
Ref.~\cite{GNC12}.

\subsection{FFCI for different polarizations and lattices:}

The FFCI state described in the main text appears for circular polarization
$A_{x}=A_{y}$. Although we have focused on $A_{x}=A_{y}=1.7$, a
similar behaviour is observed in the vicinity of such point. In this
section we show that the appearance of the FFCI state with this particular
polarization (circular) and for the honeycomb lattice is not a fine
tuned unique situation. \\

\paragraph{Elliptical polarization:}

We observe that this state also appears for elliptical polarization.
In Fig.~\ref{fig:CherSupp} (b) we show the phase diagram for the case where $A_{x}/A_{y}=\sqrt{3}$
on the $(A_{x},\phi)$ phase space. As in the circular polarization,
there are regions in the phase diagram where the effective bands have
a non trivial Chern number $c_{1}=1$. For concreteness we consider
the $S_{z}=0$ sector and $A_{x}/A_{y}=\sqrt{3}$ with $A_{y}=0.9,\phi=\pi/2$ for a $L=6\times4$
lattice with $N=8$ particles taking $\omega=10t$,
without single particle dispersion (unlike in the main text). We note that adding dispersion
only constrains the region of the phase diagram where the FFCI phase
appears but does not eliminate it from the phase diagram. As in the
main text, we compute the many-body spectrum via exact-diagonalization.
The results are shown in Fig.~\ref{fig: HC_lattice_elliptic} where
(a) shows the many body energy spectrum (b) the flux insertion and
(c) a representative local density profile.

\begin{figure}[h]
\includegraphics[scale=0.23]{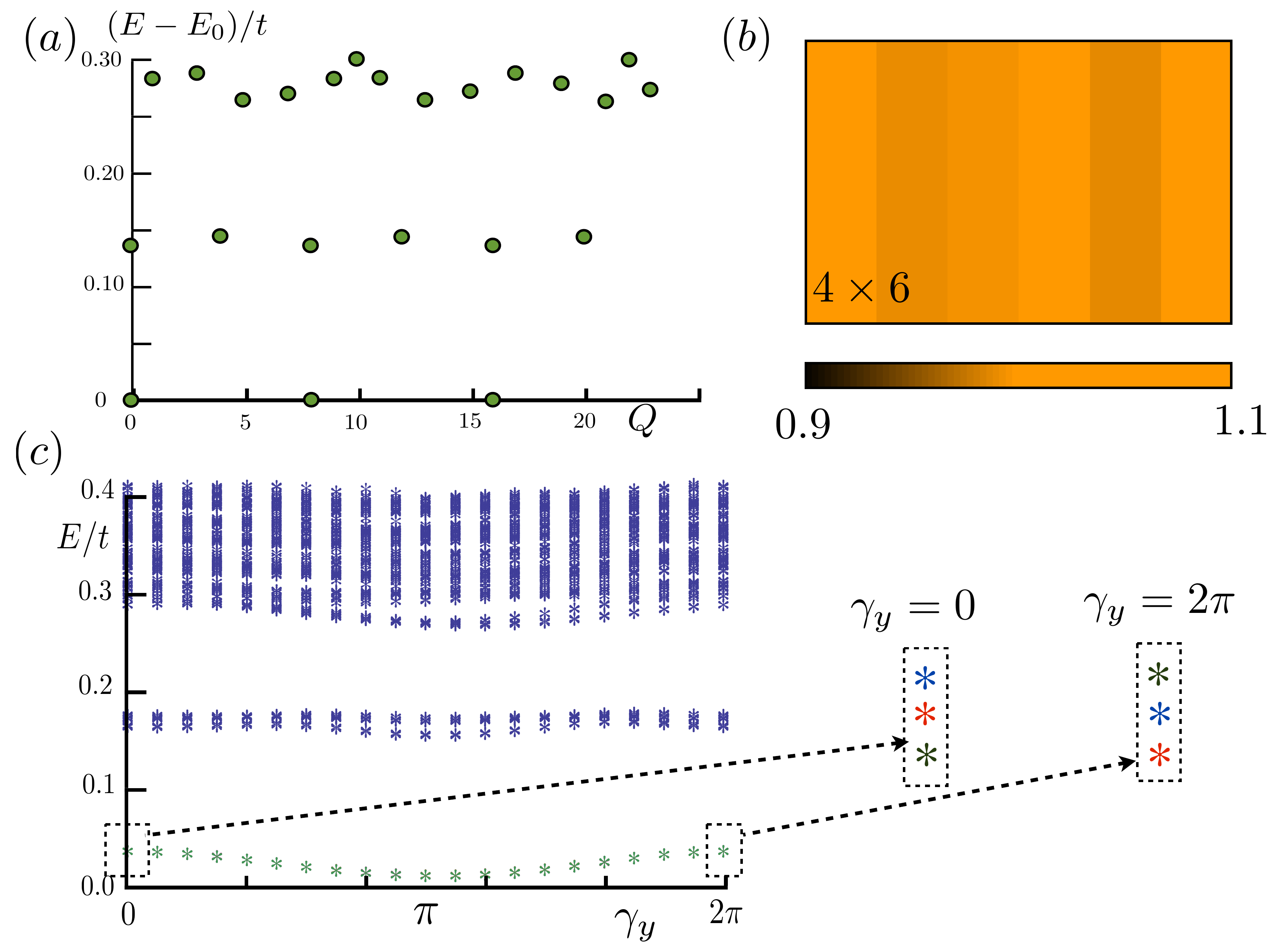} \caption{\label{fig: HC_lattice_elliptic} (a) The energy spectrum for the
driven honeycomb lattice model with $A_{x}/A_{y}=\sqrt{3}$ and $A_{y}=0.9,\phi=\pi/2$ for a
$L=6\times4$ lattice with $N=8$ particles against the total momentum
$Q$. The three FCI states appear at momenta predicted by the counting
rule of Ref.~\cite{RB11} (b) Density profile $n_{\boldsymbol{r}}^{(i)}$
defined in \eqref{eq: density n(r)} for one representative many-body
groundstate. (c) Spectral evolution of the energy spectrum
of (a) in the sector upon inserting a flux $\gamma_{y}$ into the
system. The zoom shows the interchange of the three fractional Chern
insulator groundstates under one flux period and thus their topological
degeneracy and charge fractionalization.}
\end{figure}

All three figures point to the realization of the FCI state since
the ground state fall at the total momentum sectors $Q$ expected
from the counting rule of Ref.~\cite{RB11}, they mix and interchange
with each other under adiabatic flux insertion and the density profile
is relatively uniform, as discussed in the previous section and in the main text. We nevertheless
note that the relative density change is close to $10\%$, approximately
twice the value for the circular polarization. We have also checked
that all these three features also appear for different elliptical
polarization such as $A_{x}/A_{y}=1/\sqrt{3}$.\\

\paragraph{$\pi-$flux lattice:}

The above results strongly indicate that the FFCI state described
in the main text does not depend strongly on light polarization but
rather on that the Chern number is finite. We have also studied whether
this still holds for different lattices and find that indeed the honeycomb
lattice is not a fine-tuned case. To support this claim we hereby
provide numerical evidence of the appearance of the same $1/3$ FFCI
state for the $\pi-$flux lattice at half-filling. Because of the
historical relevance of the model in condensed matter these results
are interesting on their own right and therefore here we will sum
up the most relevant results and further detail will be presented
elsewhere.\\

\begin{figure}[t]
\includegraphics[scale=0.45, page=2]{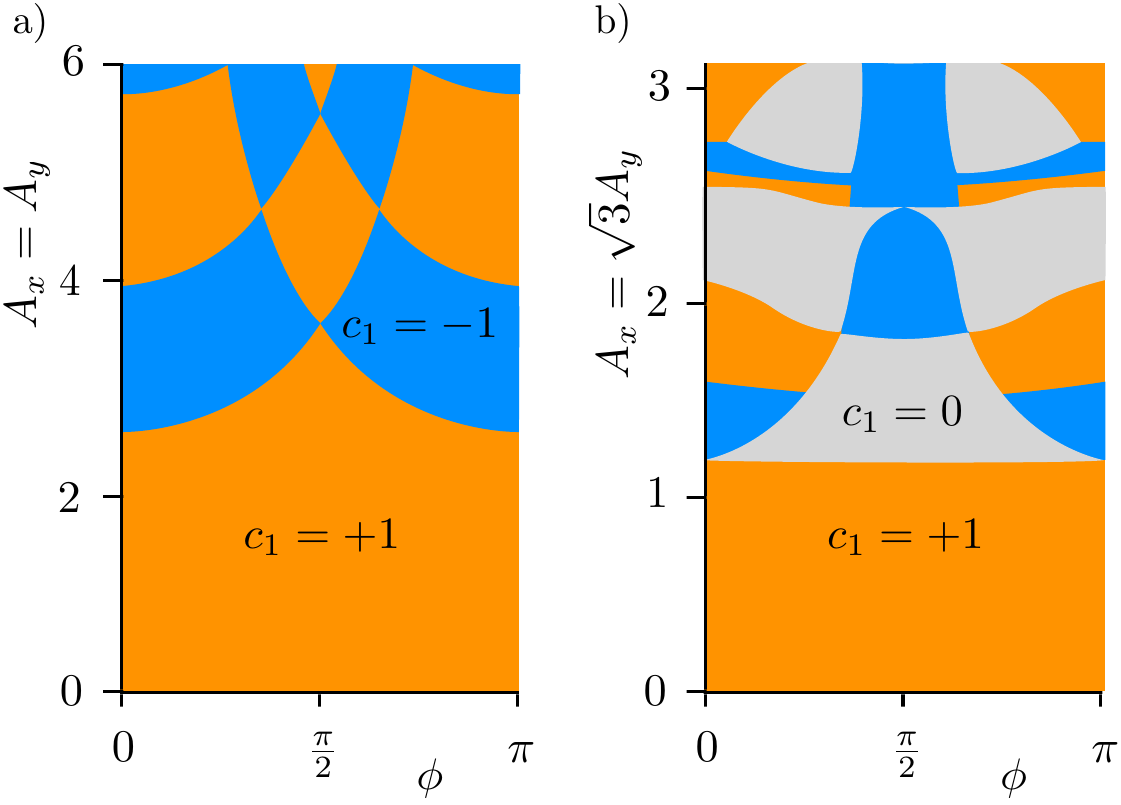} 
\caption{\label{fig:CherSupp} 
Chern number phase diagram for a) the square lattice Hamiltonian 
circularly polarized light
and b) graphene with elliptically polarized light.
}
\end{figure}

Following the procedure outlined in the first section of this supplementary
material, we have calculated the effective Hamiltonian $H_{\text{eff}}$ corresponding to the $\pi-$flux
model~\cite{WWZ89} with the conventions of Ref.~\cite{NSCM11}. The
non-interacting phase diagram for this case, analogous to Fig.~\ref{fig: Chern number phase diagram} in
the main text for the honeycomb lattice is shown in Fig.~\ref{fig:CherSupp} (a). As in
the honeycomb case, there are regions in the parameter space $(A_{x}=A_{y},\phi)$
where the Chern number of the effective bands is non-trivial and takes
the values $c_{1}=\pm1$. \\
 Again, we consider only the fully polarized spin sector $S_{z}=0$
with a $L=6\times4$ lattice with $N=8$ particles and include band
dispersion and $\omega=10t$. Choosing the particular point $A_{x}=A_{y}=1.5$
and $\phi=1.65$ corresponding to the non-trivial case with $c_{1}=1$
we find a three fold degenerate ground state as shown in the spectrum
of Fig. \ref{fig: SQ_lattice} (a) together with the flux insertion
in (b) and a representative local density profile (c). All three figures
point to the realization of the FCI state: (i) the momentum sectors
for the groundstate manifold coincide with those predicted by the
counting rule of Ref.~\cite{RB11}, (ii) the flux insertion reveals
that the groundstates evolve independently and interchange upon one
flux insertion and (iii) the uniformity of the density profile discards
the CDW state as the many body ground state. Therefore, we conclude
that the Floquet fractional Chern insulator state is also realized
in the $\pi-$flux lattice. Furthermore, the symmetries of the square
lattice make the sampling of the berry curvature BZ particularly uniform
favouring the use of the simple formula of Ref.~\cite{NSC12}. With
this we have checked that the Hall conductivity of this system is
closely quantized to $\sigma_{H}\simeq\frac{1}{3}\frac{e^{2}}{h}$
as expected for the FCI state. We found that for this lattice the
FCI state is defined in an even wider region of the phase diagram
as compared to the honeycomb case details of which will be reported
elsewhere.

\begin{figure}[h]
\includegraphics[scale=0.23]{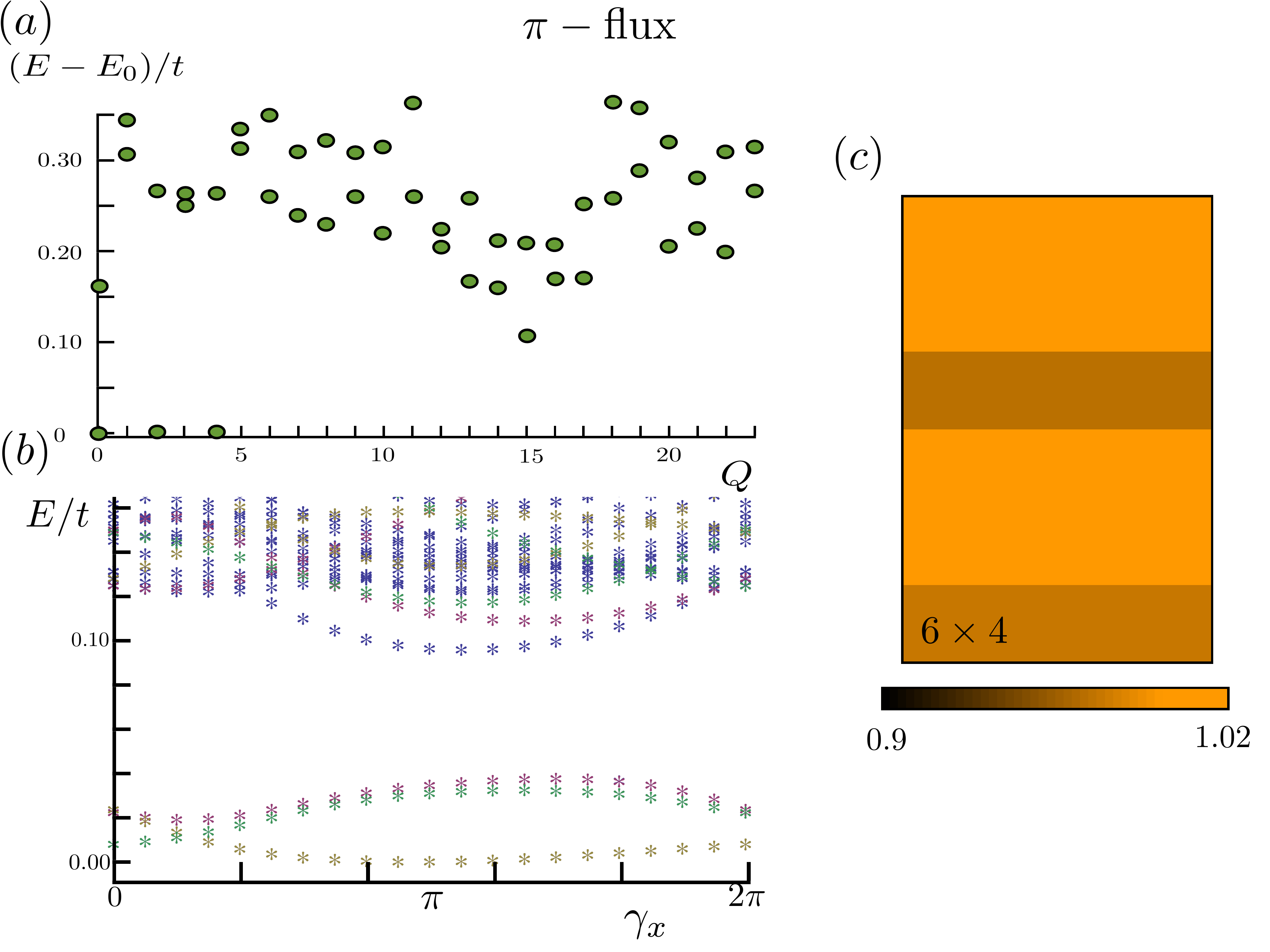} \caption{\label{fig: SQ_lattice} Floquet fractional Chern insulator state
for the $\pi-$flux model with $A_{x}=A_{y}=1.6$,$\phi=1.65$, $S_{z}=0$
and $L=6\times4$, $N=8$. (a) The energy spectrum for the $\pi-$flux
model against the total momentum $Q$. The three FCI states appear
at momenta predicted by the counting rule of Ref.~\cite{RB11} (b)
Spectral evolution of the energy spectrum of (a) in the sector upon
inserting a flux $\gamma_{x}$ into the system. The evolution of the
three fractional Chern insulator ground states signals their topological
degeneracy and charge fractionalization. c) Density profile $n_{\boldsymbol{r}}^{(i)}$
defined in \eqref{eq: density n(r)} for one representative many-body
groundstate.}
\end{figure}

\end{widetext} 
\end{document}